\documentclass[prb,amsmath,amssymb,twocolumn,showpacs,superscriptaddress]{revtex4}
\usepackage{graphicx}
\usepackage{xcolor}
\usepackage{color}
\usepackage{amsmath}
\usepackage{bm}

\definecolor{FSB-color}{named}{magenta}

\definecolor{TTH-color2}{named}{red}

\DeclareMathOperator{\sgn}{sgn}

\begin{document}

\title{Thermoelectric radiation detector based on superconductor/ferromagnet systems}

\author{T.T.~Heikkil\"a}

\author{R. Ojaj\"arvi}

\author{I. J. Maasilta}

\affiliation{Department of Physics and Nanoscience Center, University of Jyvaskyla, P.O. Box 35 (YFL), FI-40014 University of Jyv\"askyl\"a, Finland}

\author{E.~Strambini}

\author{F.~Giazotto}

\affiliation{NEST, Istituto Nanoscienze-CNR and Scuola Normale Superiore, I-56127 Pisa, Italy}

\author{F.S.~Bergeret}

\affiliation{
Centro de F\'{i}sica de Materiales (CFM-MPC), Centro Mixto CSIC-UPV/EHU, Manuel de Lardizabal 5, E-20018 San Sebasti\'{a}n, Spain}

\affiliation{Donostia International Physics Center (DIPC), Manuel de Lardizabal 5, E-20018 San Sebasti\'{a}n, Spain}

\newcommand{\fixme}[1]{\begingroup\color{red}\em(FIXME: #1)\endgroup}
\newcommand{\todo}[1]{\begingroup\color{red}\em(TODO: #1)\endgroup}
\newcommand{\tmpnote}[1]{\fixme{#1}}

\date{\today}

\begin{abstract}
We suggest a new type of an ultrasensitive detector of electromagnetic fields exploiting the giant thermoelectric effect recently found in superconductor/ferromagnet hybrid structures. Compared to other types of superconducting detectors where the detected signal is based on variations of the detector impedance, the thermoelectric detector has the advantage of requiring no external driving fields. This becomes especially relevant in multi-pixel detectors where the number of bias lines and the heating induced by them becomes an issue. We propose different material combinations to implement the detector and provide a detailed analysis of its sensitivity and speed. In particular, we perform to our knowledge the first proper noise analysis that includes the cross correlation between heat and charge current noise and thereby describes also thermoelectric detectors with a large thermoelectric figure of merit. 
\end{abstract}

\maketitle

\section{Introduction}

Some of the most accurate sensors of wide-band electromagnetic radiation are based on superconducting films. Such sensors, in particular the transition edge sensor (TES), are used in a wide variety of applications requiring extremely high sensitivity. Those applications include detection of the cosmic microwave background \cite{cmbref,cmbref2,cmbref3}  and other areas of astrophysics \cite{farrahup17}, generic-purpose terahertz radiation sensing used for example in security imaging \cite{luukanen12}, gamma-ray spectroscopy of nuclear materials \cite{bennett} and materials analysis via detection of fluorescent x-rays  excited by ion beams \cite{palosaari2016}, short laser-driven x-ray pulses \cite{miaja-avila} or syncrotrons \cite{bennett}. Many of these applications would benefit from adding more pixels, i.e., more sensors, to improve the collection efficiency, detection bandwidth or the spatial or angular resolution. However, operating large arrays of TES sensors can become problematic  as each pixel requires a bias line. This can become cumbersome in the presence of thousands of pixels, even with advanced multiplexing techniques \cite{bennett}. Moreover, the bias lines tend to carry heat into the system espacially by radiation, reducing for example its overall noise performance. In addition, TES always dissipates power at the pixel, giving constraints on the cryogenic design for large arrays. One alternative is the kinetic inductance detector (KID) \cite{kidref} and its variants \cite{pjs}, the most common being the type with passive frequency-domain multiplexing using superconducting microwave resonators \cite{day}. With such a device, a single pair of coaxial cables can be used to probe a large array of pixels, but the probe power is by necessity also partially dissipated at the detectors.  

Both TES and KID sensors are based on the measurement of an impedance of the sensor, i.e., response to a probe signal. It would generally be beneficial if one could get rid of the probe signal altogether, so that the measured signal would result directly from the radiation coupled to the detector. This is what happens in thermoelectric detection \cite{jones47,varpula17, van1999thermoelectric}, where the temperature rise caused by the absorption of radiation is converted into an electric  voltage or current that can then be detected. Such thermoelectric detectors have been discussed before, but they have not been considered for ultrasensitive low-temperature detectors for the simple reason that thermoelectric effects are typically extremely weak at low temperatures. On the other hand, at high temperatures where such thermoelectric effects would be strong enough, the thermal noise hampers the device sensitivity.

We suggest to overcome these problems in a {\it superconductor-ferromagnet thermoelectric detector} (SFTED) \cite{patent} by exploiting the newly discovered giant thermoelectric effect taking place in superconductor/ferromagnet heterostructures \cite{ozaeta14,machon13,kolenda16,bergeret17} for radiation sensing. As this thermoelectric effect can be realized with close to Carnot efficiency \cite{ozaeta14,bergeret17} even at sub-Kelvin temperatures, the resulting detector can have a large signal-to-noise ratio, and a noise equivalent power (NEP) rivaling those of the best TES and KID detectors without the burden of having to use additional bias lines for probing the sensor,  and with zero (for ideal amplification) or at most very small non-signal power dissipation at the sensor location. The only part of the system where external power is needed is in the detection of the thermoelectric currents, i.e. at the amplifier, which
can be taken far from the active sensing region.

\begin{figure}[h]
\centering
\includegraphics[width=\columnwidth]{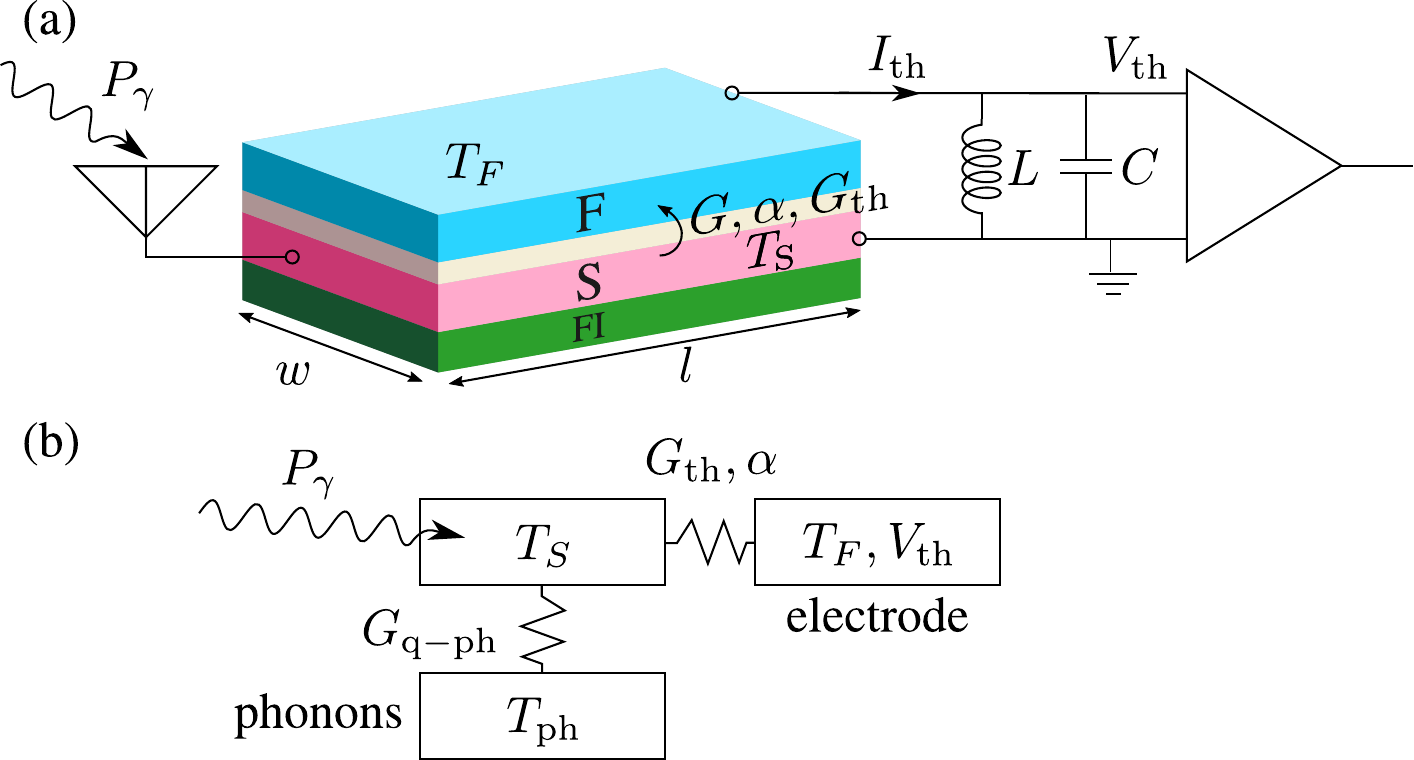}
\caption{(a) Schematic of the thermoelectric detector, where a temperature difference $T_S-T_F$ drives a thermoelectric current $I_{\rm th}$ and/or thermovoltage $V_{\rm th}$ across a spin-polarized junction. The latter is either composed of a normal insulator and a ferromagnetic electrode, or a ferromagnetic insulator and a normal metal electrode. (b) Heat balance: incoming radiation power heats up the quasiparticles in the spin-split superconductor $S$, and the amount of heating depends on the heat conductances to the main heat baths.}
\label{fig:SFTS}
\end{figure}

Recently the use of SF structures in thermometry has been discussed \cite{giazotto15}. Despite some similarities, thermometers and radiation detectors have quite different requirements regarding their sensitivity. In particular, the sensitivity of the radiation detectors typically is dictated by the temperature fluctuation noise, which is not an issue as such for thermometers.  In this paper we concentrate on exactly finding and optimizing the relevant figures of merit for radiation detection, and hence cannot benefit much from the results in thermometry.

For concreteness, we consider the detector realization depicted in Fig.~\ref{fig:SFTS}. The sensor element, i.e., one pixel of a possible detector array, is formed from a thin film superconductor-ferromagnetic insulator (S-FI) bilayer coupled to superconducting (S') antennas via a clean (Andreev) contact. This bilayer is further connected, via a tunnel junction (magnetic or normal) to a ferromagnetic electrode F \cite{FInote}. The current injected to the ferromagnetic electrode or the voltage generated across the tunnel junction is detected by a SQUID current amplifier or a field effect transistor, respectively. In what follows, we describe conditions for measuring radiation in the sub-mm/far-IR regime, in which case it can be coupled to the detector via antennas. Alternatively, the detector could be used for measuring radiation at higher frequencies (such as X-rays) in which case the system should be connected to an additional larger absorber element.

We consider the radiation to be directed to the detector via a superconducting antenna (not in the figure) which is coupled via a clean (Andreev) contact to the active S-FI region (absorber). To prevent heat leaking out from the absorber, the superconductor used in the antenna should be fabricated from a material with a higher superconducting gap $\Delta_A$ than the one used in the absorber, $\Delta$. One possible combination could be Nb antenna and an Al absorber. For optimal quantum efficiency, the normal state resistance of the absorber (seen by the radiation at frequencies higher than $\Delta/h$, where $h$ is Planck's constant) should be matched to the specific impedance of the antenna, typically somewhat below the vacuum impedance. For Al film thickness of 10 nm, a typical sheet resistance is 5\dots 10 $\Omega/\Box$.  Hence, a 1 $\mu$m wide film with length $l=10$ $\mu$m would have the resistance $R_\gamma = $ $50 \dots 100$ $\Omega$ seen by the radiation, thereby matching well with typical antennas. In what follows, we hence choose an absorber region of this size. Reducing (increasing) the width and length while keeping their ratio constant would result to the same quantum efficiency, but decreased (increased) noise and dynamic range.

The absorber superconductor (S) is placed in contact with a ferromagnetic insulator (FI) that exerts a magnetic proximity effect on the former, resulting into a spin-splitting exchange field $h$ inside S. In Al, large induced spin-splitting fields have been detected by contacting it for example with EuS \cite{moodera88} or EuO \cite{tedrow86}. At low temperatures compared to the S critical temperature $T_C$, the exchange field does not have a major effect on the order parameter $\Delta$ \cite{clogston62,alexander86}. However, it results into a strong (and opposite) electron-hole asymmetry in each spin component in the direction specified by the FI magnetization. This asymmetry can be used to generate a thermoelectric signal if S is connected via a spin filter to another electrode \cite{ozaeta14,bergeret17}. This spin filtering is provided here by the ferromagnetic electrode $F$ and is quantified by the normal-state spin polarization $P=(G_\uparrow-G_\downarrow)/(G_\uparrow +G_\downarrow) \in [-1,1]$, where $G_\sigma$ is the normal-state conductance of the S-F contact for spin channel $\sigma$. In what follows, we also characterize this contact via its spin-averaged normal-state conductance $G_T$. In practice, oxide contacts with ferromagnetic metals such as Ni, Co or Fe have $P \sim 0.1 \dots 0.45$ \cite{meservey94} whereas using ferromagnetic insulator contacts  may lead to polarizations exceeding $P \sim 0.9$ \cite{moodera07}. The precise value of $G_T$ for a given area of the junction can be controlled with the thickness of the tunnel junction. 

\section{Noise equivalent power of a thermoelectric detector}

Let us first consider a generic thermoelectric element working as a radiation sensor and analyze its figures of merit, in particular the noise equivalent power NEP and the thermal time constant $\tau_T$, both defined in detail below. As the radiation with power $P_\gamma$ is absorbed in the absorber, it first creates a strong nonequilibrium state of the quasiparticles. This nonequilibrium state relaxes via (i) quasiparticle-quasiparticle (q-q) collisions, via (ii) spurious processes such as the quasiparticle-phonon (q-ph) relaxation, and (iii) via the escape of the quasiparticles to the (ferromagnetic) electrode. The last process yields the detected signal. Moreover, (iv) some of the excitations may escape as quasiparticles to the antenna. We assume that the process (i) dominates so that the quasiparticles thermalize between themselves before escaping to the antenna, and therefore in what follows we disregard process (iv). As a result of this chain of events, the quasiparticles in the absorber heat up to the temperature $T_S=T+\Delta T$ determined from a heat balance equation \cite{currentnote} 
\begin{equation}
C_h \frac{d \Delta T}{dt} = P_\gamma - G_{\rm th}^{\rm tot}\Delta T + \alpha V_{\rm th},
\end{equation}
where $C_h$ is the heat capacity of the absorber, $G_{\rm th}^{\rm tot}=G_{\rm q-ph}+G_{\rm th}$, and $G_{\rm q-ph}$ and $G_{\rm th}$ denote the heat conductances from quasiparticles to the phonons and to the ferromagnetic electrode, respectively. In the linear regime we assume both to reside at the bath temperature $T$. The last term  results from the Peltier heat current driven by the induced thermovoltage across the S-F junction, and it is also proportional to the temperature difference $\Delta T$. We assume the detector to operate at low powers $P_\gamma$ so that these linear response relations are sufficient.  The detector characteristics depends strongly on the chosen $T$.

The induced temperature difference (in frequency domain)  $\Delta T=(P_\gamma+\alpha V_{\rm th})/(i\omega C_h + G_{\rm q-ph}+G_{\rm th})$  drives a thermoelectric current $I_{\rm th}=\alpha \Delta T/T-GV_{\rm th}$ into the ferromagnet and ultimately to an amplifier. To focus on detector performance limits first, we disregard the back-action noise from the amplifier, and consider the amplifier only as a reactive element; either a capacitor or an inductor, corresponding to the field effect transistor or SQUID amplifier, respectively. Therefore, the thermoelectric current equals $V_{\rm th} [i\omega C +1/(i \omega L)]$ across the amplifier with capacitance $C$ and inductance $L$ in parallel.  The practical limits of voltage (current) measurements can be obtained by considering $\omega \neq 0$ and taking the limit $L\rightarrow \infty$ ($C\rightarrow \infty$).  From these relations we can obtain the voltage and current responsivities,
\begin{equation}
\lambda_V\equiv \frac{V_{\rm th}}{P_\gamma} = \frac{\alpha}{Y_{\rm th}^{\rm tot} Y^{\rm tot}T-\alpha^2}, \quad \lambda_I \equiv \frac{I_L}{P_\gamma} = \frac{\lambda_V}{i\omega L},
\end{equation}
where $I_L$ is the current across the inductor. The relevant responsivity depends on the choice of the amplifier. Here $Y_{\rm th}^{\rm tot}=i\omega C_h + G_{\rm th}^{\rm tot}$ and $Y^{\rm tot}=G+i\omega C + 1/(i \omega L)$ are the thermal and electrical admittances, respectively. Note that $ZT(\omega)=\alpha \lambda_V$ is a finite-frequency generalization of the usual thermoelectric figure of merit. 

Let us then consider the temperature fluctuation $\delta T$, voltage noise $\Delta V$ across the capacitor and the current noise $\Delta I_L$ across the inductor. These are driven by the three intrinsic noise sources: the charge and heat current noises $\delta I$ and $\delta \dot Q_J$ across the thermoelectric junction and the heat current noise $\delta \dot Q_{\rm q-ph}$ for the quasiparticle-phonon process. Now the heat balance equation and Kirchoff law for the noise terms read
\begin{subequations}
\begin{align}
Y_{\rm th}^{\rm tot} \delta T &= \delta \dot Q_{\rm q-ph} + \delta \dot Q_J + \alpha \Delta V\\
Y^{\rm tot} \Delta V &= \delta I + \alpha \delta T/T.
\end{align}
\end{subequations}
Solving these yields 
\begin{equation}
\Delta V = \lambda_V (\delta \dot Q_{\rm q-ph}+\delta \dot Q_J+Y_{\rm th}^{\rm tot} T \delta I/\alpha)
\end{equation}
and $\Delta I_L=\Delta V/(i\omega L)$.  To find the second-order correlator of these noise terms, we assume that the intrinsic correlators satisfy
\begin{subequations}
\begin{align}
\langle \delta I^2 \rangle &= 4 k_B T G\\
\langle \delta \dot Q_J^2 \rangle &= 4 k_B T^2 G_{\rm th}\\
\langle \delta I \delta \dot Q_J \rangle &= -4 k_B T \alpha\label{eq:crossnoise}\\
\langle \delta \dot Q_{\rm q-ph}^2 \rangle &= 4 k_B T^2 G_{\rm q-ph}.
\end{align}
\end{subequations}
These result from the fluctuation-dissipation theorem for the individual contacts. In particular, the cross-noise term is important for strong thermoelectric response, and was not taken into account before, it was for example disregarded in \cite{varpula17}. The total voltage noise spectral density is (note that this is the symmetrized voltage noise correlator, and therefore one needs to take the absolute value squared)
\begin{equation}
S_V = \langle \Delta V^2 \rangle = |\lambda_V|^2 4 k_B T^2 \frac{G_{\rm th}^{\rm tot}(GT G_{\rm th}^{\rm tot}-\alpha^2)+\omega^2 C_h G T}{\alpha^2},
\label{eq:voltnoise}
\end{equation}
The term in parenthesis is positive semidefinite due to the thermoelectric stability condition $\alpha^2 \le G T G_{\rm th}^{\rm tot}$ valid for all thermoelectric systems. The current noise spectral density across the inductor, $S_I$, has the same form as $S_V$ in \eqref{eq:voltnoise}, but where $\lambda_V$ is replaced by $\lambda_I$. 

The noise equivalent power squared ($NEP^2$) is the power spectral density for which the induced thermoelectric voltage spectral density across the capacitor equals $S_V$, or the thermoelectric current spectral density across the inductor equals $S_I$. These yield the same results, $S_V/|\lambda_V|^2=S_I/|\lambda_I|^2$, 
\begin{equation}
NEP^2 \equiv \frac{S_V}{|\lambda_V|^2} = 4 k_B T^2 \frac{G_{\rm th}^{\rm tot} (GT G_{\rm th}^{\rm tot}-\alpha^2) + \omega^2 C_h^2 GT}{\alpha^2}.
\end{equation}
This may be written in a more tractable form by using the  zero-frequency thermoelectric figure of merit $zT=\alpha^2/(G_{\rm th}^{\rm tot} G T-\alpha^2)$ \cite{ZTnote} and the thermal time constant $\tau_T=C_h/G_{\rm th}^{\rm tot}$,
\begin{equation}
NEP^2 = \frac{4 k_B T^2 [1+\omega^2 \tau_T^2 (1+zT)] G_{\rm th}^{\rm tot}}{zT}.
\label{eq:NEPtot}
\end{equation}
 The zero-frequency thermoelectric $NEP$ thus equals the usual thermal bolometer NEP \cite{giazotto06} from the thermal fluctuation noise, divided by the square root of the figure of merit. Moreover, the thermal time constant determining the frequency band for the detection is increased by the factor $\sqrt{1+zT}$. 
 
 In the above discussion we have disregarded the contribution from the amplifier noise. This is discussed separately below.

The NEP written above can be optimized, as typically the overall level of the thermoelectric junction conductance can be chosen almost at will, and the coefficients $\alpha$ and $G_{\rm th}$ scale with the same prefactor. In the sensor discussed here, this means optimizing the normal-state conductance of the thermoelectric junction. On the other hand, for a given absorber volume $\Omega$, the thermal conductance of the spurious process $G_{\rm q-ph}$ cannot be affected much. Therefore, choosing a too small junction conductance results in a poor thermoelectric figure of merit $zT$, whereas increasing the junction conductance increases the thermal fluctuation noise and the Johnson-Nyquist current noise of the junction. In addition, a high junction conductance may lead to a heating of the normal-metal (ferromagnetic) electrode, thereby reducing the temperature gradient across the junction, and the associated thermoelectric effects. In what follows we assume that this electrode is thick enough so that its heating can be disregarded. For zero-frequency NEP, the optimum is obtained with a conductance $G_{\rm th}/G_{\rm ph}=\sqrt{1+zT_i}$, where $zT_i=\alpha^2/(G_{\rm th} G T-\alpha^2)$ is the intrinsic figure of merit of the junction (not including the heat conductance to the phonons). With this choice, the optimal NEP of the thermoelectric detector is 
\begin{equation}
NEP^2_{\rm opt} = \frac{4 G_{\rm q-ph} k_B T^2 (1+\sqrt{1+zT_i})^2}{zT_i}.
\end{equation}
It hence reaches the limit set by the thermal fluctuation noise of the spurious process for $zT_i \rightarrow \infty$. 

The above discussion holds for arbitrary thermoelectric sensors of the type depicted in Fig.~\ref{fig:SFTS}b. However, typically strong thermoelectric effects are found only above room temperature, which renders the thermal fluctuation noise very large. The combination of a spin-split superconductor with a spin-polarized contact circumvents this problem, leading to large thermoelectric response even at sub-Kelvin temperatures. In the following, we analyze this system in more detail. 

\section{Superconductor/ferromagnet thermoelectric radiation detector}

First, following \cite{ozaeta14}, we write the thermoelectric coefficients of the spin-polarized junction between the spin-split superconductor and the non-superconducting contact as
\begin{subequations}
  \label{eq:numcoefs}
  \begin{align}
    G&=G_T \int_{-\infty}^\infty dE \frac{N_0(E)}{4 k_B T \cosh^2\left(\frac{E}{2k_B T}\right)}
    \,,
    \\
    G_{\rm th}&=\frac{G_T}{e^2} \int_{-\infty}^\infty dE \frac{E^2 N_0(E)}{4 k_B T^2 \cosh^2\left(\frac{E}{2k_B T}\right)}
    \,,
    \\
    \alpha&=\frac{G_T}{2e} P \int_{-\infty}^\infty dE \frac{E N_z(E)}{4 k_B T \cosh^2\left(\frac{E}{2k_B T}\right)}
    \,.
\end{align}
\end{subequations}
Here $G_T$ is the normal-state electrical conductance of the junction, $N_0(E)=(N_\uparrow+N_\downarrow)/2$ and $N_z=N_\uparrow-N_\downarrow$ are the spin-averaged and the spin-difference density of states (DOS) of the superconductor, normalized to the normal-state DOS $\nu_F$ at the Fermi level. They are obtained from $N_{\uparrow/\downarrow} = N_S(E\mp h)$ with $N_S(E)={\rm Re}[|E+i\Gamma|/\sqrt{(E+i\Gamma) ^2-\Delta^2}]$, where $h$ is the spin-splitting field and $\Gamma \ll \Delta$ describes pair-breaking effects inside the superconductor. Analytic approximations for Eqs.~\eqref{eq:numcoefs} are detailed in \cite{ozaeta14}.

The heat capacity of the absorber with volume $\Omega$ is obtained from
\begin{equation}
\begin{split}
C_h =& \frac{d}{dT} \left\{\nu_F \Omega \int_{-\infty}^{\infty} dE E N_0(E) f_{\rm eq}(E)\right\} \\=& \frac{\nu_F}{4 k_B T^2} \int_{-\infty}^{\infty} \frac{E^2 N_0(E)}{\cosh^2 \left(\frac{E}{2k_B T}\right)} = \frac{\nu_F \Omega e^2}{G_T} G_{\rm th}.
\end{split}
\end{equation}
The thermal time constant of the junction, $C_h/G_{\rm th}$, hence remains independent of superconductivity or spin splitting.

The electron-phonon heat conductance of a spin-split superconductor in the pure limit can be obtained from\cite{virtanen16}
\begin{equation}
\begin{split}
G_{\rm e-ph} = \frac{\Sigma \Omega}{96 \zeta(5) k_B^5 T^2}&
\int_{-\infty}^\infty dE E \int_{-\infty}^\infty d\omega \\&\times\omega^2
|\omega| L_{E,E+\omega} F_{E,\omega},
\end{split}
\label{eq:gqpph}
\end{equation}
with 
\begin{equation}
\!F_{E,\omega}{=}-\frac 1 2\!\left[\sinh\!\left(\frac{\omega}{2T}\right)\cosh\!\left(\frac{E}{2T}\right)\cosh\!\left(\frac{E{+}\omega}{2T}\right)\right]^{-1},
\end{equation}
$L_{E,E'}=\frac{1}{2}\sum_{\sigma=\pm} N_\sigma(E) N_\sigma(E') \{1-\Delta^2/[(E+\sigma h)(E'+\sigma h)]\}$ and $\sigma=\pm$ for spin $\uparrow/\downarrow$. Here $\Sigma$ is the materials dependent electron-phonon coupling constant (for typical values, see \cite{giazotto06}), $\Omega=wld$ is the volume of the S island, and $\zeta(x)$ is the Riemann zeta function. 

In the low-temperature limit \(k_B T\ll\Delta-|h|\), the electron-phonon heat conductance can be approximated as
\begin{equation}
\begin{split}\label{eq:Gep_approx}
G_{\rm e-ph} = \frac{\Sigma\Omega}{96\zeta(5)}T^4 \big[\cosh \tilde h\,e^{-\tilde\Delta} &f_1(\tilde\Delta)\\
+ \pi\tilde\Delta^5e^{-2\tilde\Delta} &f_2(\tilde\Delta) \big].
\end{split}
\end{equation}
where \(\tilde h=h/k_B T\) and \(\tilde\Delta =\Delta/k_B T\). The function $f_1$ can be approximated with an expansion \(f_1(\tilde\Delta) = \sum_{n=1}^\infty C_n/\tilde \Delta^n\) with coefficients \(C_0\approx 440\), \(C_1\approx -500\), \(C_2\approx 1400\), \(C_3\approx -4700\). An expansion for \(f_2\) is \(f_2(\tilde\Delta) = \sum_{n=1}^\infty B_n/\tilde \Delta^n\) with coefficients \(B_0 = 64\), \(B_0 = 64\), \(B_1 = 144\) and \(B_2 = 258\). The derivation of Eq.~\eqref{eq:Gep_approx} is presented in the appendix.

\begin{figure}[h]
\centering
\includegraphics[width=\columnwidth]{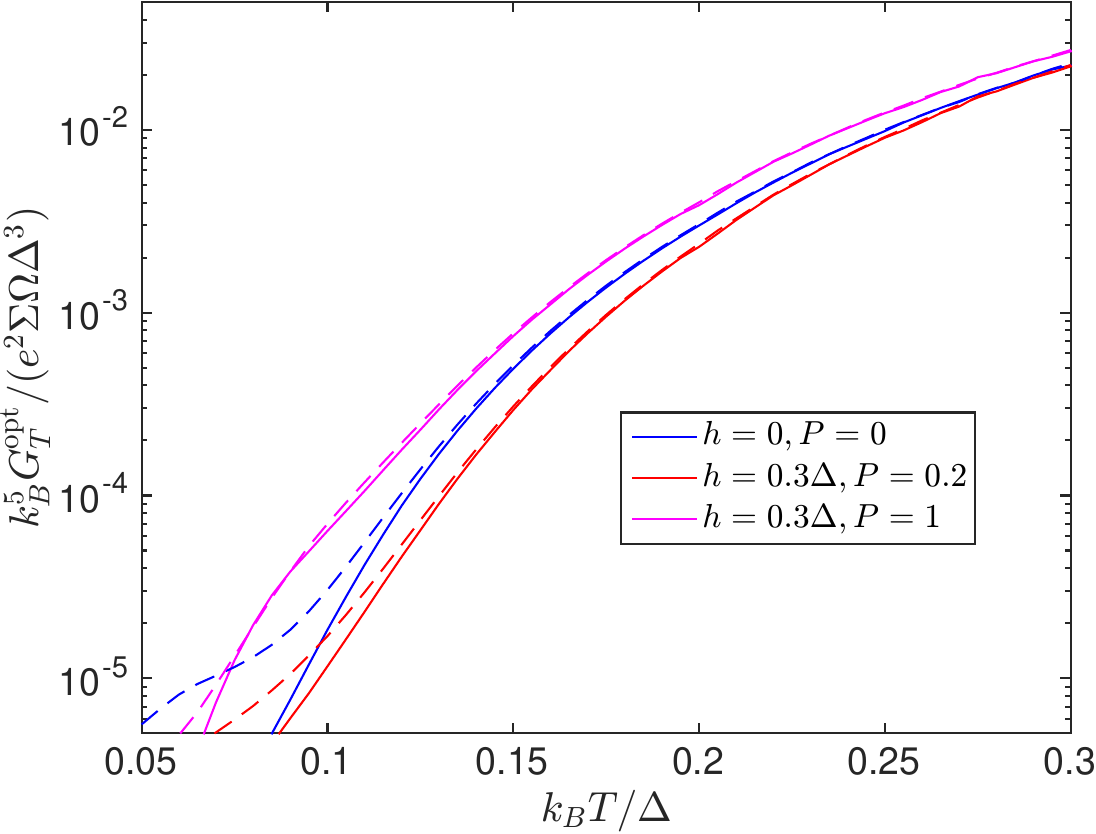}
\caption{Optimal normal-state tunnel conductance $G_T$ of the thermoelectric junction. For Al of volume $10^{-19}$ m$^3$, the value of one corresponds to $G_T^{-1}=20$ $\Omega$. At the lowest temperatures the results start to depend on the chosen broadening parameter $\Gamma$ used in the numerics. The solid lines were calculated with $\Gamma=10^{-4} \Delta$ and the dashed lines with $\Gamma=10^{-3} \Delta$.}
\label{fig:optimalG}
\end{figure}

In the following, we employ the above formulas to discuss the behavior of the SFTED. For this, we evaluate the above integrals numerically to obtain predictions of the optimal junction conductance, the thermoelectric figure of merit $zT$, the total NEP and the time constant. The optimal normal-state junction conductance $G_T$ is plotted as a function of temperature in Fig.~\ref{fig:optimalG}. As the electron-phonon heat conductance becomes relatively weaker than the junction conductance at low temperatures, the optimal junction conductance also depends strongly on temperature. Note that for an Al absorber with electron-phonon coupling constant $\Sigma=0.3$ $\times 10^9$ W/(m$^3$ K$^5$), volume $\Omega=10^{-19}$ m$^3$ and $\Delta=200$ $\mu$eV, the dimensionless parameter $k_B^5 G_T/(e^2 \Sigma \Omega \Delta^3)=G_T \times 20$ $\Omega$. The optimal normal-state resistance of the junction is thus within the range 20 k$\Omega$ $\dots$ 20 M$\Omega$. Moreover, since both $G_T$ and $\Sigma \Omega$ depend on the area of the absorber, the real optimizable parameters are the absorber film thickness and the junction conductance per area. In what follows, we use $G_T=5\times 10^{-4} e^2 \Sigma \Omega \Delta^3$ corresponding to a junction resistance of 40 k$\Omega$, optimal roughly at $T \approx 0.1 \Delta/k_B \sim 200$ mK. This corresponds to a resistance times unit area of 400 k$\Omega \, \mu$m$^2$, which is quite easily reached with AlO$_2$ tunnel junctions \cite{strambini17}, but would be somewhat challenging for  
spin-filter EuS barriers, ranging typically between 10-1000 M$\Omega \, \mu$m$^2$ \cite{moodera88}.

\begin{figure}[h]
\centering
\includegraphics[width=\columnwidth]{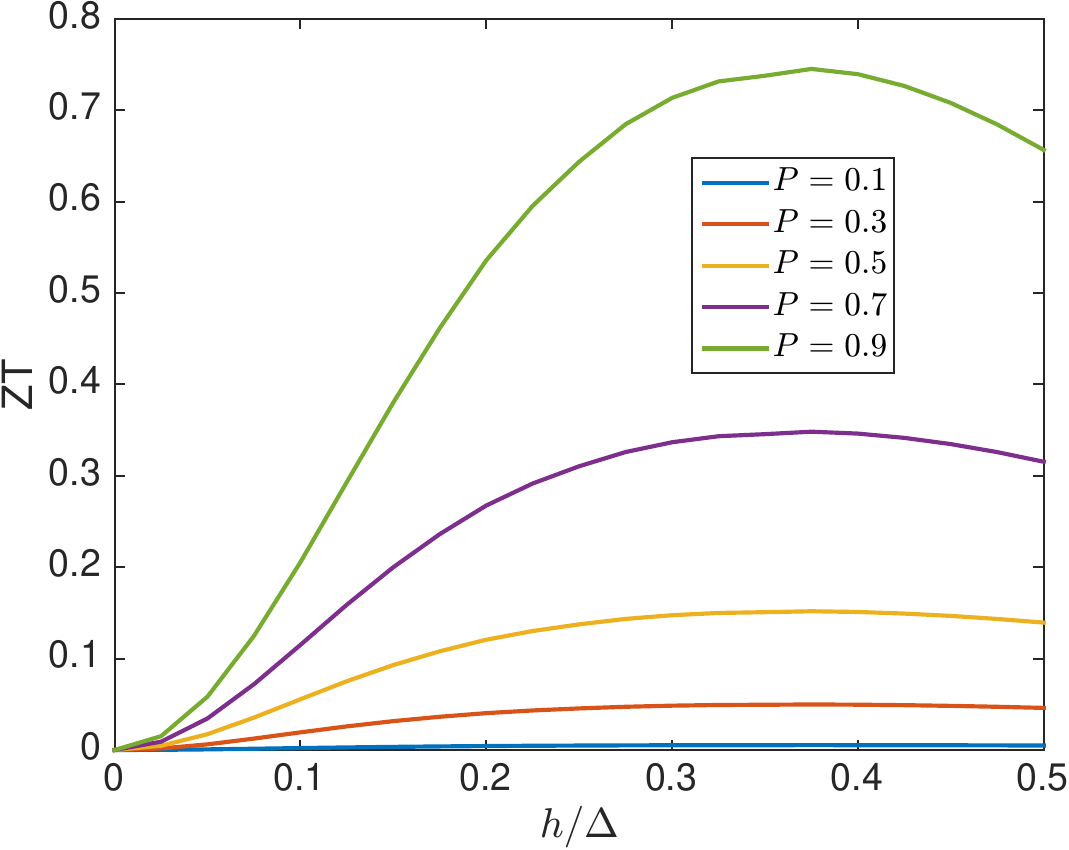}
\caption{Thermoelectric figure of merit as a function of the exchange field for junctions with different polarizations $P$, at the temperature $T=0.1\Delta/k_B$, with $G_T=5\times 10^{-4} e^2 \Sigma \Omega \Delta^3$ and $\Gamma=10^{-3} \Delta$.}
\label{fig:ZT}
\end{figure}

On the other hand, the thermoelectric figure of merit $zT$ depends strongly on the detector polarization. We show this by plotting $zT$ for the parameters indicated above as a function of the exchange field at $T=0.1\Delta/k_B$ in Fig.~\ref{fig:ZT}. Due to the presence of the electron-phonon process acting as an extra heat channel, the figure of merit does not exceed unity. 

\begin{figure}[h]
\centering
\includegraphics[width=\columnwidth]{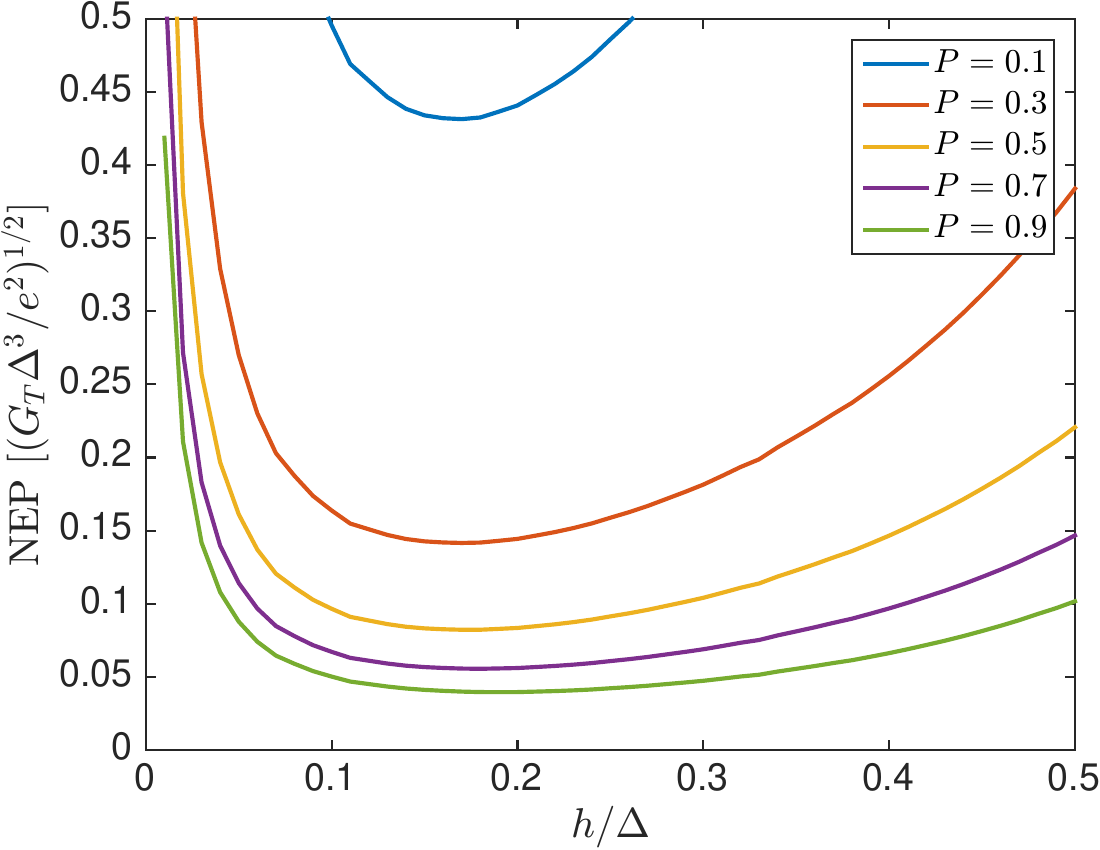}
\caption{Zero frequency noise equivalent power as a function of the exchange field for junctions with different polarizations $P$, at the temperature $T=0.1\Delta/k_B$, with $G_T=5\times 10^{-4} e^2 \Sigma \Omega \Delta^3$ and $\Gamma=10^{-3} \Delta$. For the parameters considered in this paper, $\sqrt{G_T \Delta^3/e^2} \approx 10^{-18}$ W/$\sqrt{\rm Hz}$.}
\label{fig:NEPvsh}
\end{figure}

\begin{figure}[h]
\centering
\includegraphics[width=\columnwidth]{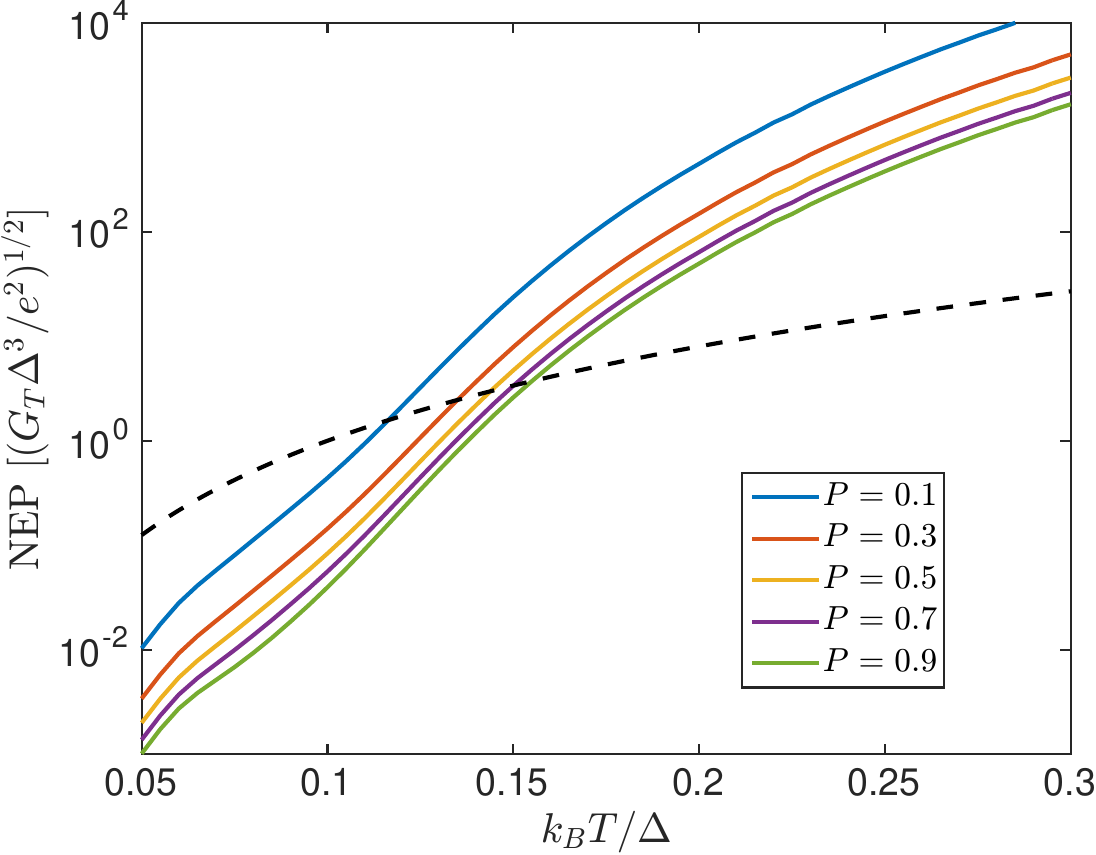}
\caption{Zero frequency noise equivalent power as a function of the temperature for junctions with different polarizations $P$, with the exchange field $h=0.2\Delta$, with $G_T=5\times 10^{-4} e^2 \Sigma \Omega \Delta^3$ and $\Gamma=10^{-3} \Delta$. For the parameters considered in this paper, $\sqrt{G_T \Delta^3/e^2} \approx 10^{-18}$ W/$\sqrt{\rm Hz}$. The dashed line shows the thermal fluctuation noise NEP=$\sqrt{20 \Sigma \Omega T^6}$ for a transition edge sensor of the same volume.}
\label{fig:NEPvsT}
\end{figure}

The most interesting characteristic of any detector is its sensitivity, in this case the noise equivalent power. This we plot as a function of exchange field in Fig.~\ref{fig:NEPvsh} and temperature in Fig.~\ref{fig:NEPvsT}. The NEP value is normalized to $\sqrt{G_T \Delta^3/e^2}$, which corresponds to approximatively $10^{-18}$ W/$\sqrt{\rm Hz}$ for the chosen parameters. The dashed lines indicate the NEP$_{\rm bolo}=4 G_{\rm q-ph}T^2$  obtained for a transition edge sensor TES with the same absorber volume at the corresponding temperature, with its heat conductance limited by electron-phonon coupling, i.e., a hot-electron TES \cite{karasik11}. In Fig.~\ref{fig:NEPvsh} that reference value happens to be exactly unity for the chosen parameters. As TES operates in the dissipative regime at the transition, the normal state value for $G_{\rm q-ph}$ must be used. Moreover, this estimate disregards the bias-induced heating, which sets the operating temperature higher than the bath temperature, and extra noise sources often found in TES realizations. We find that SFTED can reach similar or better values than such a TES even with quite modest values of the junction polarization at low temperatures. Note that these results depend a bit on the precise value of the junction conductance --- with a higher conductance, NEP at higher temperatures would be lower (see Fig.~\ref{fig:optimalG}), and vice versa. 

In practice, the most sensitive TES bolometers up to date have been fabricated from suspended structures where the thermal conductance to the bath is limited by phonon transport, achieving $NEP$ values of the order of $1 \times 10^{-19}$ W/$\sqrt{\rm Hz}$ \cite{beyer12,suzuki16} at $T_C$ around 100 mK. Based on Fig. 5, the SFTED device is also predicted to reach a lower NEP than that.

\begin{figure}[h]
\centering
\includegraphics[width=\columnwidth]{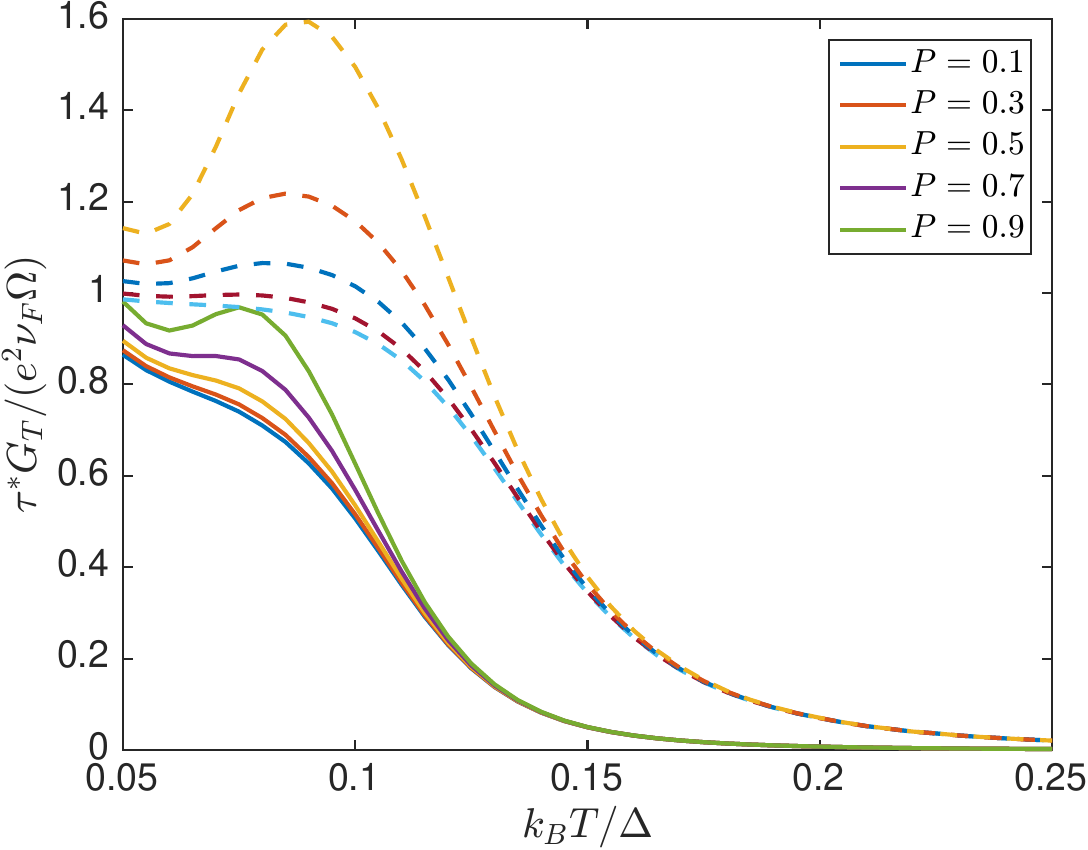}
\caption{Temperature dependence of the detector time constant $\tau^*$ determining the (angular) frequency bandwidth $1/\tau^*$ where the NEP is low. We have chosen $h=0.2\Delta$. Solid lines are calculated with $G_T=5\times 10^{-4} e^2 \Sigma \Omega \Delta^3$, whereas  the dashed lines correspond to 10 times larger conductance. For the previous, $\tau^0 \approx 0.1$ ms, and for the latter it is 0.01 ms.}
\label{fig:timeconstant}
\end{figure}

For completeness, we show the behavior of the thermal time constant $\tau^*=\tau_T \sqrt{1+zT}$ as a function of temperature in Fig.~\ref{fig:timeconstant}. It is given in units of $\tau_0=\nu_F \Omega e^2/G_T$. For $\nu_F=10^{47}$ 1/(J m$^3$), $\Omega=10^{-19}$ m$^3$, and $G_T^{-1}=2$ M$\Omega$, $\tau_0 \approx 0.1$ ms. At low temperatures, the tunnel junction dominates the heat conductance, and $\tau^* \approx \tau_0$. In this case $zT$ is also appreciable, and slightly modifies $\tau^*$. On the other hand, at high temperatures electron-phonon heat conduction takes over, and the detector becomes faster. To illustrate this crossover, we show the time constant for two different values of $G_T$. 

The above results were obtained by disregarding spin relaxation. Aluminum is a light material, and therefore the spin-orbit scattering in it is typically quite weak, and the spin relaxation is dominated by spin-flip scattering. The typical spin relaxation times $\tau_{sn}$ in Al  are of the order of 100 ps, \cite{poli08} and therefore $\hbar/(\tau_{sn}\Delta) \sim 0.03$, and the model disregarding spin relaxation is more or less justified. However, spin-flip scattering in the presence of exchange field yields a non-zero density of states inside the superconducting gap, and eventually leads to pair breaking \cite{bergeret17}. Above, such effects are taken  into account with the parameter $\Gamma$. For heavier materials, such as Nb, spin relaxation is caused by spin-orbit scattering, and the thermoelectric effects become weaker. Therefore, using such heavier materials for example to increase the operation temperature of the thermoelectric detector beyond the critical temperature of Al would require further analysis of the effects of spin relaxation.  

\subsection{Contribution of amplifier noise}
The above analysis has been made by disregarding the noise due to the voltage or current measurement. We can include it by assuming an added voltage noise spectral density $S_V^A$ or current noise spectral density $S_I^A$ for the amplifier used for voltage or current measurement. In the case of voltage measurements the added an amplifier NEP contribution is (for $\omega=0$, for simplicity)
\begin{equation}
NEP_{\rm A,V}^2 = \frac{S_V^A G T G_{\rm th}^{\rm tot}}{zT(1+zT)},
\end{equation}
whereas in the case of current measurement the contribution is
\begin{equation}
NEP_{\rm A,I}^2 = \frac{S_I^A (1+zT) G_{\rm th}^{\rm tot}T}{G zT}.
\end{equation}
We can hence see that the relative contribution from the voltage amplifier to the overall NEP decreases as the thermoelectric junction resistance increases. On the other hand, in the case of current measurement the amplifier contribution becomes independent of the junction resistance when $G_{\rm th}^{\rm tot}$ is dominated by the junction heat conductance. Another way to estimate the contribution of amplifier noise is by dividing the corresponding NEP values by the total thermoelectric NEP from Eq.~\eqref{eq:NEPtot} (at $\omega=0$). We hence get
\begin{subequations}
\begin{align}
r_V\equiv \frac{NEP_{\rm A,V}^2}{NEP^2} &= \frac{S_V^A}{4 k_B T}\frac{G}{(1+zT)}\\
r_I\equiv \frac{NEP_{\rm A,I}^2}{NEP^2} &= \frac{S_I^A }{4 k_B T } \frac{(1+zT)}{G} .
\end{align}
\end{subequations}
A typical good voltage preamplifier for low-frequency measurements has a voltage noise of the order of $\sqrt{S_V} = $ 1.5 nV$/\sqrt{{\rm Hz}}$ at room temperature and $\sqrt{S_V} = $ 0.3 nV$/\sqrt{{\rm Hz}}$ for cryogenic amplifiers \cite{beev13}. 
Combining this value with the normal-state tunnel conductance and the $\Delta$ chosen above for Al, the relative NEP for voltage measurement is $r_V \approx G \Delta/(G_T k_B T) (1+zT)^{-1}$. 
This is much below unity in the entire relevant temperature range ($k_B T \ll \Delta$) due to the exponential suppression of $G$. On the other hand, a very good current amplifier can have an added noise of $\sqrt{S_I} = $ 0.5 fA$/\sqrt{{\rm Hz}}$.
With that value we get $r_I \approx 10^{-4} \Delta/(k_B T) \times  (1+zT) G_T/G $.
This exceeds unity below $k_B T \approx 0.1 \Delta$ (precise value depending on the chosen exchange field), and the current measurement accuracy starts limiting the TED NEP below those temperatures. This difference between the two types of measurements originates from the fact that the thermoelectric voltage can be of the order of the temperature difference itself due to the thermopower of the order of $k_B/e$, whereas the thermoelectric current is exponentially suppressed \cite{ozaeta14} and hence harder to measure. However, note that ultimately at very low temperatures  the voltage measurement also becomes harder as it requires the voltmeter impedance to far exceed that of the junction, and this condition becomes harder to meet at low temperatures.

Typical absolute thermometer -based radiation detectors operating at the bath temperature (i.e., in contrast to for example TES where the bias sets the operating point above bath temperature) suffer from chip temperature fluctuations due to fluctuations in the cooling power. However, a thermoelectric detector measures a temperature difference $\Delta T$ instead of the absolute temperature. Because the chip temperature fluctuations affect both the temperature of the absorber and that of the measurement electrode, they do not affect $\Delta T$ (to the lowest order). This is an added benefit for TEDs in comparison with detectors based on resistance or inductance measurements. 

\section{Conclusions}

\begin{figure}[h]
\centering
\includegraphics[width=\columnwidth]{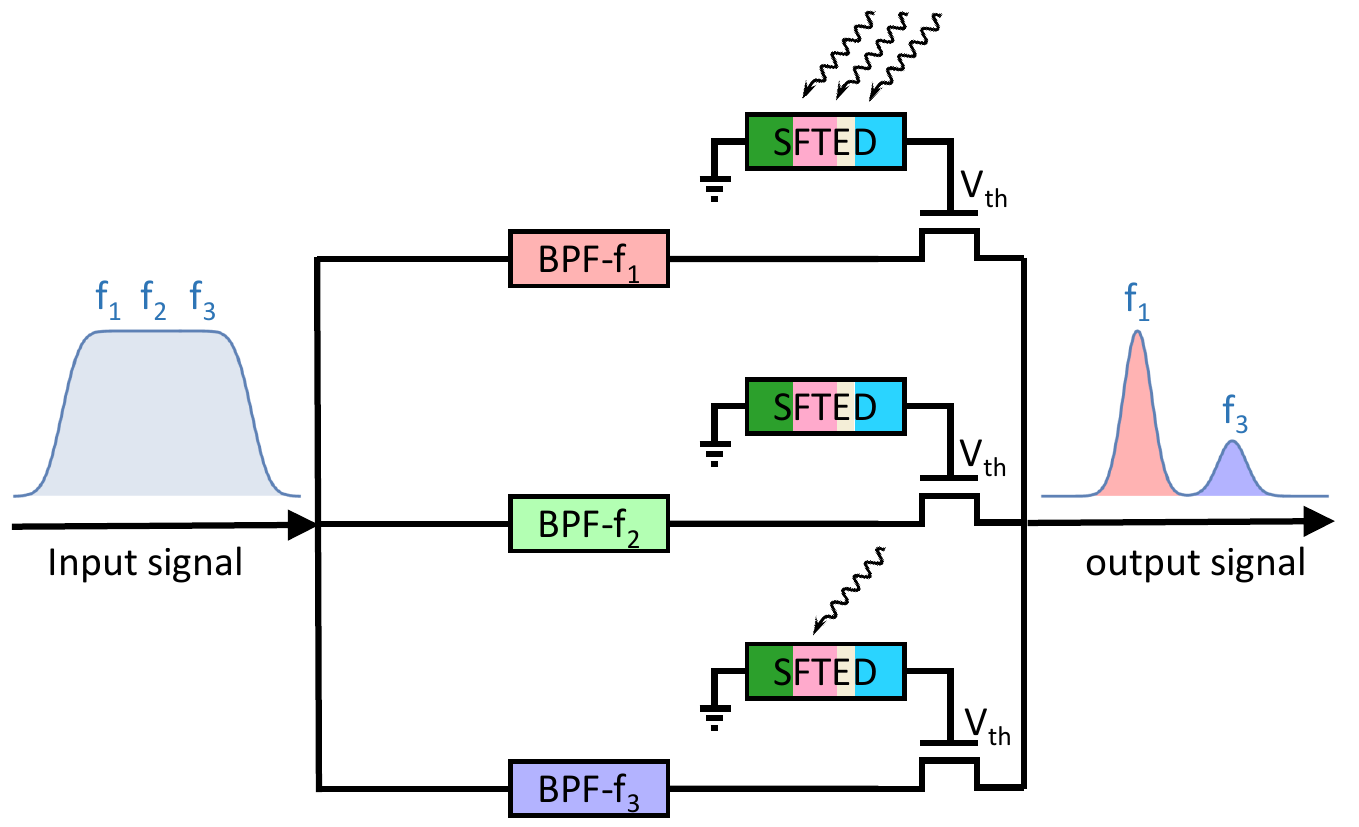}
\caption{One possible scheme for frequency-domain multiplexing of the thermoelectric detectors. Here a broad-band electromagnetic wave, or a frequency comb, is divided into different frequency components via narrow band-pass filters (BPF), and directed through field-effect transistors whose conductance is modulated by the voltage from the detectors. As a result, the output spectrum contains pixel-specific information about the absorbed radiation power.}
\label{fig:multiplex}
\end{figure}

As such thermoelectric radiation detection is not a new concept \cite{jones47}.  However, most of the previously studied thermoelectric detectors have relied on using semiconducting thermoelectric materials, operating at and above room temperature $T_{\rm RT}$.  Because the spurious heat conduction processes have a heat conductance scaling at least as $\sim T^3$ (typical phonon heat conductivity \cite{giazotto06}), the corresponding NEP is $(T_{\rm RT}/T )^{5/2}  \sim 10^{15/2}$ times larger than that considered here (this estimate assumes the Debye temperature to exceed the room temperature, but it should in any case be taken as indicative).  On the other hand, quantum dot structures may exhibit strong thermoelectric effects even at low temperatures \cite{svilans16}.  Contrary to the superconductor/ferromagnet structure considered here, in those devices the thermoelectric effects are single-channel phenomena, and therefore it may be difficult to make the  electronic thermal conduction dominate over the spurious heat conduction channels.

In this paper, we have shown how  a combination of superconducting and magnetic materials can be used to construct a truly novel type of  a low-temperature radiation detector relying on the thermoelectric effect and thereby not requiring extra bias power to be applied into the device. This leads to simpler designs of arrays of such detectors, and helps in maintaining the  low operating temperatures required for ultrasensitive operation. In addition, ultrasensitive TES bolometers necessarily have a very low tolerance for excess power loading, as the device can be saturated and pushed out of the transition region with it. For the detectors discussed here there is no such abrupt effect, although excess power could lead to performance degradation due to overheating. Nevertheless, due to the lack of the bias lines, novel multiplexing strategies may need to be  designed. We present one possible scheme in Fig.~\ref{fig:multiplex}. There, the output looks quite similar to that of frequency multiplexed TES or KID readout schemes, but the possible heating effects in the (dissipative) field-effect transistors can be engineered far apart from the pixels absorbing the radiation. Nevertheless, the optimal multiplexing strategies is a topic for further research.

\acknowledgments

We thank Alexander Stefanescu, Bruno Leone and Subrata Chakraborty for discussions. This project was supported by Academy of Finland Key Funding project (Project No. 305256), the Center of Excellence program (Project No. 284594) and project 298667. F.G. aknowledges the European Research Council under the European Union’s Seventh Framework Program (FP7/2007-2013)/ERC Grant agreement No. 615187-COMANCHE, and Tuscany Region under the FARFAS 2014 project SCIADRO
for partial financial support. 
The work of F.S.B. was supported by Spanish Ministerio de Economia, Industria  y Competitividad (MINEICO) through Projects No. FIS2014-55987-P and FIS2017-82804-P.

\appendix

\section{Electron-phonon heat conductance}

Let us calculate the electron-phonon heat conductance coefficient \(G_{\rm e-ph}\) beginning from Eq. (12) in the main text.
We first extract the temperature dependent prefactor by scaling all the quantities with dimensions of energy by temperature. For example, \(\tilde E = E/k_B T\). The scaled quantities are dimensionless and are denoted with a tilde over the variable.  We then change the integration variables to \(x=\tilde E\) and \(y=\tilde E+\tilde\omega\).  We obtain
\begin{equation}\label{eq:GepI}
G_{\rm e-ph} = \frac{\Sigma\Omega T^4}{96\zeta(5)} \sum_{\sigma=\pm}\frac{I^\sigma}{4},
\end{equation}
where
\begin{equation}\label{eq:splitintegral}
\begin{split}
I^\sigma &= \iint\!dx dy \sgn[(x+\sigma \tilde h)(y+\sigma \tilde h)] \times \\
&\frac{x |x-y|^3\left[(x+\sigma \tilde h)(y+\sigma \tilde h) - \tilde \Delta^2\right]}{\sqrt{ ([x+\sigma \tilde h]^2-\tilde\Delta^2 )( [y+\sigma \tilde h]^2 - \tilde\Delta^2 )}}\frac{4 e^{-\frac{|x|+|y|}{2}}}{\sinh\frac{x-y}{2}}.
\end{split}
\end{equation}
Above, both $x$ and $y$ are integrated from \(-\infty\) to \(+\infty\), excluding the region \([-\tilde\Delta-\sigma \tilde h,\tilde\Delta-\sigma \tilde h]\) in which the spin-split DOS vanishes. We also assumed that \(\Delta - h \gg k_B T\) so that we could make an approximation
\begin{equation}\label{eq:coshapprox}
\cosh x \approx \frac{e^{|x|}}{2}, \quad x>\tilde\Delta-\tilde h,
\end{equation}
and similarly for $\cosh y$.

The integral is divided into four separate quadrants by the gaps in the DOS. The integral over the quadrant $n$ for the spin $\sigma$ is $I^\sigma_n$. Because the integrand of Eq.~\eqref{eq:splitintegral} is symmetric with respect to simultaneous inversion of $x$, $y$ and $\sigma$, the contributions from the opposing quadrants are equal,
\begin{equation}\label{eq:quadrants}
I = \frac 1 4 \sum_{n=1}^4 \sum_{\sigma=\pm} I^\sigma_n = \frac 1 2 \sum_{\sigma=\pm}\left( I^\sigma_1 + I^\sigma_2 \right).
\end{equation}

Let us calculate the integral over the first quadrant. This part of the integral represents scattering processes, for which we have in the earlier variables \(E>0\) and \(E'>0\). Thus, the interacting quasiparticles are both particle-like. By shifting the integration limits, we get
\begin{align}
I^\sigma_1 &= \int_{\tilde\Delta-\sigma\tilde h}^\infty \mkern-18mu dx \int_{\tilde\Delta-\sigma\tilde h}^\infty \mkern-18mu dy \frac{x |x-y|^3\left[(x+\sigma h)(y+\sigma h) - \tilde \Delta^2\right]}{\sqrt{ ([x+\sigma h]^2-\tilde\Delta^2 )( [y+\sigma h]^2 - \tilde\Delta^2 )}}\nonumber\\
&\quad\quad\times \frac{8}{e^x - e^y} \nonumber\\
&= \int_0^\infty\mkern-10mu dx \int_0^\infty\mkern-10mu dy \frac{(x + {\tilde{\Delta}}-\sigma \tilde h) |x-y|^3 \left[xy + (x+y){\tilde{\Delta}}\right]}{(e^{x} - e^{y})\sqrt{x y (x+ 2 {\tilde{\Delta}}) (y+ 2 {\tilde{\Delta}})}}\nonumber\\
&\quad\quad\times 8 e^{-{\tilde{\Delta}+\sigma\tilde h} }.
\end{align}
Above, we have $\tilde h$-dependence in two places, in the exponential outside the integral and as a linear term in the numerator. However, the parts of the numerator which are symmetric with respect to exchange \(x\leftrightarrow y\) do not contribute to the integral. Therefore, we can write the integral as
\begin{align}
I^\sigma_1 &= e^{-{\tilde{\Delta}+\sigma\tilde h}} \int_0^\infty\mkern-15mu dx \int_0^\infty\mkern-15mu dy \frac{8 x^2 |x-y|^3 (y + \tilde\Delta )}{(e^{x} - e^{y})\sqrt{x y (x+ 2 {\tilde{\Delta}}) (y+ 2 {\tilde{\Delta}})}}\nonumber\\
& = e^{-{\tilde{\Delta}+\sigma\tilde h}} f_1(\tilde\Delta),
\end{align}
where $f_1(\tilde\Delta)$ is a monotonically increasing function with values $f(2)\approx 326$ and $\lim_{\tilde\Delta\to\infty} f_1(\tilde\Delta) \approx 438$. A Taylor expansion $f_1(\tilde\Delta) = \sum_{n=0}^\infty C_n/{\tilde\Delta^n}$ can be calculated by first expanding the integrand into series in $\tilde\Delta^{-1}$ and then doing the integral separately for each term. The values of the first few coefficients are $C_0 \approx 440$, $C_1 \approx -500$, $C_2 \approx 1400$, $C_3 \approx -4700$.

Doing the sum over the spins, we find the contribution from the first quadrant,
\begin{equation}\label{eq:quadrant1}
I_1 = \sum_{\sigma=\pm} I^\sigma_1 = 2 \cosh \tilde h e^{-\tilde\Delta} f_1(\tilde\Delta).
\end{equation}

The second quadrant describes the contribution from the recombination processes, for which one quasiparticle is hole-like (\(E<0\)) and the other particle-like (\(E'>0\)).
\begin{align}
I_2^\sigma = &-\int_{-\infty}^{-\tilde\Delta-\sigma \tilde h}\mkern-15mu dx \int_{\tilde \Delta-\sigma \tilde h}^\infty\mkern-15mu dy\; \frac{8 e^x}{e^x-e^y}\\
&\quad\quad\times \frac{x (x-y)^3\left[(x+\sigma h)(y+\sigma h) - \tilde \Delta^2\right]}{\sqrt{ ([x+\sigma h]^2-\tilde\Delta^2 )( [y+\sigma h]^2 - \tilde\Delta^2 )}}\nonumber\\
=& \int_0^\infty\mkern-10mu dx \int_0^\infty\mkern-10mu dy \frac{(x+y+2 {\tilde{\Delta}})^3 \left[xy +(x+y){\tilde{\Delta}} +2{\tilde{\Delta}}^2\right]}{\sqrt{x y (2 {\tilde{\Delta}} +x) (2 {\tilde{\Delta}} +y)}}\nonumber\\
&\quad\quad\times 8(x+{\tilde{\Delta}} +\sigma\tilde h) e^{-x-y}e^{-2 {\tilde{\Delta}}}.\nonumber
\end{align}
where we approximated \(e^{2 \Delta +x+y}-1\approx e^{2 \Delta +x+y}\). Above, the exchange field $\sigma \tilde h$ appears only as a linear term. Summing over the two spin directions, terms odd in $\sigma$ cancel and we can write $I_2$ in the form 
\begin{equation}\label{eq:quadrant2}
I_2 = \sum_{\sigma} I^\sigma_2 = 2\pi\tilde\Delta^5 e^{-2\tilde \Delta} f_2(\tilde\Delta).
\end{equation}
Within the approximation \eqref{eq:coshapprox}, the exchange field does not modify the contribution from the recombination processes.

The function $f_2$ is defined as
\begin{align}
f_2(\tilde\Delta) = \int_0^\infty\mkern-10mu dx \int_0^\infty\mkern-10mu dy &\frac{(x{+}y{+}2 {\tilde{\Delta}})^3 \left[xy +(x{+}y){\tilde{\Delta}} +2{\tilde{\Delta}}^2\right]}{\pi\tilde\Delta^5 \sqrt{x y (2 {\tilde{\Delta}} +x) (2 {\tilde{\Delta}} +y)}}
\nonumber\\
\times\; &8e^{-x-y}(x+{\tilde{\Delta}})
\end{align}
The function $f_2$ is a monotonically decreasing function with values \(f_2(4)\approx 123\) and \(\lim_{\tilde\Delta\rightarrow\infty}f_2(\tilde\Delta) = 64\). An expansion $f_2(\tilde\Delta) = \sum_{n=0}^\infty B_n/{\tilde\Delta^n}$ is obtained by first expanding the integrand asymptotically at $\tilde\Delta=\infty$ and then calculating the integral term by term. The values of the first few coefficients are $B_0=64, B_1=144, B_2=258$ and $B_3=693/2$.

By combining Eqs.~\eqref{eq:GepI}, \eqref{eq:quadrants}, \eqref{eq:quadrant1} and \eqref{eq:quadrant2}, we find the electron-phonon heat conductance for a spin-split superconductor, Eq.~\eqref{eq:Gep_approx}. 
At low temperatures, when $\tilde\Delta \gg 1$, scattering processes dominate the heat conductance. The two processes become of the same order of magnitude when $k_B T\approx 0.1 \Delta$. At high temperatures, recombination processes dominate.


\begin{thebibliography}{99}
\bibitem{FInote} The system presented in Fig.~\ref{fig:SFTS} is perhaps the simplest realization that does not require an external magnetic field. Alternatively, the upper S-I-F part could be replaced by either S-FI-N or S-FI-S, where the ferromagnetic insulator FI would act as the spin filter. In that case it would also be possible to remove the lower FI.
\bibitem{currentnote} We choose the convention where the heat currents are "into" the absorber, and charge currents are "out of" it. Alternative conventions lead to some sign changes in the equations.
\bibitem{ZTnote} Note that $zT$ is not the zero-frequency limit of $ZT$, but should be considered as a definition.
\end{thebibliography}
\end{document}